\documentclass[11pt]{article}
\usepackage{latexsym}

\usepackage{amsthm}
\usepackage{amsfonts}
\usepackage{amssymb}
\newcommand {\N}{\mathbb{N}} 
\newcommand {\Z}{\mathbb{Z}} 
\newcommand {\R}{\mathbb{R}} 
\newcommand {\PP}{\mathbb{P}} 
\newcommand {\E}{\mathbb{E}} 

\newtheorem{thm}{Theorem}[section]
\newtheorem{lem}[thm]{Lemma}

\newtheorem{de}[thm]{Definition}

\def\eps{\varepsilon}

\def\bd{\begin{displaystyle}}
\def\ed{\end{displaystyle}}

\def\fin{\hfill $\Box$}

\def\eps{\varepsilon}

\def\phin{\varphi_{n,\omega}}
\def\x{x_{n,\omega}}
\def\En{E_{n,\omega}}

\begin{document}
\title{Dynamical Localization for the Random Dimer Schr\"odinger Operator}
\author{Stephan De Bi\`evre \\e-mail: debievre@gat.univ-lille1.fr\\
\and Fran\c cois Germinet\\e-mail: germinet@gat.univ-lille1.fr
\and \and
UFR de Math\'ematiques et URA GAT\\Universit\'e
des Sciences et Technologies de Lille\\ 59655 Villeneuve d'Ascq Cedex\\France}
\date{1999}
\maketitle

\begin{abstract} We study the one-dimensional random dimer model, with
Hamiltonian
$H_\omega=\Delta + V_\omega$, where for all $x\in\Z, V_\omega(2x)=V_\omega(2x+1)$
and where the $V_\omega(2x)$ are i.i.d. Bernoulli random variables taking the
values $\pm V,\; V>0$.  We show that, for  all values of $V$ and with
probability one in $\omega$, the spectrum of $H$ is pure point. If $V\leq1$ and $V\neq 
1/\sqrt{2}$, the Lyapounov exponent vanishes only at the two critical energies given by 
$E=\pm V$. For the particular value $V=1/\sqrt{2}$, respectively $V=\sqrt{2}$, we show the 
existence of additional critical energies at $E=\pm 3/\sqrt{2}$, resp. $E=0$. On any compact 
interval $I$
not containing the critical energies, the eigenfunctions are then shown to be
semi-uniformly exponentially localized, and this implies dynamical localization:
for all $q>0$ and for all
$\psi\in\ell^2(\Z)$  with sufficiently rapid decrease:
$$
\sup_t r^{(q)}_{\psi,I}(t) \equiv \sup_t \langle P_I(H_\omega)\psi_t, \ |X|^q
P_I(H_\omega)\psi_t \rangle\ <\infty.
$$ Here $\psi_t=e^{-iH_\omega t} \psi$, and $P_I(H_\omega)$ is the spectral
projector of $H_\omega$ onto the interval $I$. In particular if $V>1$ and $V\neq 
\sqrt{2}$, these results hold on
the entire spectrum (so that one can take $I=\sigma(H_\omega)$).
\end{abstract}

\section{Introduction}

We study a one-dimensional discrete Schr\"odinger operator, known as the  random
dimer model, introduced in \cite{dunlap}. 
 More precisely, the family of Hamiltonians $H_\omega$
($\omega\in\Omega=\{0,1\}^{\Z}$) that we consider is defined as follows. For $u\in
\ell^2(\Z)$,
\begin{equation} (H_\omega u)(x)=u(x-1)+u(x+1)+V_\omega(x)u(x),\; x\in\Z,
\label{eq:dimer}
\end{equation} where $V_\omega(2x+1)=V_\omega(2x)$, and the
$(V_\omega(2x))_{x\in\Z}$ are independent and identically distributed random
variables, with $\PP(V_\omega(0)=-V)=p$, $0<p<1$ and $V>0$. Note that the on-site
potential takes only two values and takes the same value on pairs of sites,
whence the name of the model which has attracted considerable attention in the
physics  literature
 since it seems to display an interesting localization-delocalization phenomenon 
\cite{dunlap} \cite{evec} \cite{wgp} that we now briefly explain. 

When $V\leq1$, it is easy to see that, due to a resonance  phenomenon, there is
perfect transmission at two critical energies $E_c=\pm V$. In other words, at
these energies, the model has a delocalized eigenstate \cite{fh}. It is then argued in
\cite{dunlap} that, when considering the model  constrained to a box of size
$N$,  the inverse  localization length (Lyapounov exponent) of the
eigenfunctions behaves as
$\gamma(E)\sim |E-E_c|^2$ (a result confirmed by a perturbative calculation in
\cite{bov} \cite{fh}), such that roughly $\sqrt N$ of the $N$ eigenfunctions have a
localization length of the order of the size of the box.  Using these
observations on the eigenfunctions, the authors of \cite{dunlap} argue that
$\langle\psi_t, X^2
\psi_t\rangle$ behaves like $t^{3/2}$ when $\psi_0$ is a state initially 
localized at the origin, a result they confirm with numerical computations.  In
other words, according to those results, the random dimer model is a simple
model in which a diverging localization length at isolated energies in the band
could lead to  superdiffusive behaviour.

This conclusion has been be contested on several grounds. It is argued in
\cite{gangsen} that the  behaviour in $t^{3/2}$ is only a transient effect, that
would disappear if  one explored $\langle\psi_t, X^2 \psi_t\rangle$ numerically
over much longer times than was done in \cite{dunlap}. Their objections are
essentially based on the way the $N\to\infty$ and $t\to \infty$ limits are taken
in \cite{dunlap}, and on the observation that the fraction of delocalized states
over localized states behaves as $1/\sqrt N$, so that the role of the
delocalized states may vanish in the infinite lattice model. This latter
argument is
 already proposed in
\cite{lifgrepas}, in the context of other, similar models. 

Without settling the question of the $t^{3/2}$ behaviour, we provide in this
letter some rigorous results on the random dimer model that should help to clarify
the situation. First, one expects  that in the infinite model, whatever the
value of $V$,  the Hamiltonian has pure point spectrum with  exponentially
localized eigenfunctions. Second,   when $V>1$, the
$t^{3/2}$-behaviour should be completely suppressed in the sense that 
$\sup_t\langle\psi_t, X^2\psi_t\rangle<\infty$, a property we refer to as
``dynamical localization".  This is is indeed proven in
 Theorem \ref{thprim} ($V\neq \sqrt{2}$).

It is furthermore agreed on by
all authors that, in the case $V<1$,
 the  superdiffusive behaviour -- if any -- can only come from contributions of
the eigenstates close to the critical energies. We give a precise content to
this statement and a proof of it in Theorem \ref{thdimer}.

 To obtain these results, we proceed as follows.  We first show that for all
energies $E$ away from the critical energies, the corresponding eigenfunctions
are semi-uniformly exponentially localized (this notion is introduced in \cite{iv}), {\em 
i.e.}:
$$
|\psi_E(x)| \leq C_\eps\exp |x_E|^{\eps} \exp -\gamma_E|x-x_E|,
$$
with $\eps>0$, where $x_E$ is a point where $\psi_E$ reaches its maximum and
$\gamma_E$ is the (strictly positive) Lyapounov exponent.  This, together with
the results of  \cite{dynloc} implies in turn dynamical localization.
This result has been announced in \cite{dynloccras}.

We insist once again that our results do not imply  the absence of the
superdiffusive behaviour
observed by \cite{dunlap} when the disorder is low ($V<1$): we actually feel
this model should indeed display such behaviour, but to prove it requires 
lower
bounds on the eigenfunctions close to the critical energies, rather than the
above upper bounds. It would  be interesting, since it would provide a random model
with pure point  spectrum in which $\langle\psi_t, X^2\psi_t\rangle$ has a
non-trivial lower bound at all times $t$.

We also exhibit the existence of new critical energies (in the sense that the Lyapunov 
exponent vanishes) for the special values $V=1/\sqrt{2}$ and $V=\sqrt{2}$. This is the 
content of Theorem~\ref{thcrit}. To our opinion, the nature of
these energies is different from the one of $E=\pm V$, and
should not lead to a delocalization phenomenon, but we did not prove this (see section 
\ref{sectcrit} for more details).

%
%
\section{Theorems and Localization}
\label{dimer}

We first rewrite the eigenvalue equation $H_\omega u=E\,u$  as follows:
\[
\left( \begin{array}{c} u(x+1) \\ u(x) \end{array} \right)
  = S^E_{V_\omega(x)} \left( \begin{array}{c} u(x) \\ u(x-1) \end{array} \right),
\mbox{ where } S^E_v =
                   \left( \begin{array}{cc}
                              E-v &-1 \\
                              1   &0
                          \end{array}
                   \right),
\] is the usual one-step transfer matrix. In the present case the structure of
the potential leads us to consider the two-step random transfer matrices 
$T^E_v=(S^E_v)^2$, {\em i.e.}:

\[ T^E_v=\left( \begin{array}{cc}
                (E-v)^2-1 & -(E-v) \\
                (E-v)     & -1
             \end{array}
      \right).
\]

\begin{de}
We'll say that $H_\omega$, as in (\ref{eq:dimer}), is dynamically localized on a
spectral interval $I$, iff with
probability one, for all $q>0$ and for all exponentially decaying initial state $\psi\in 
\ell^2(\Z)$:
$$
\sup_t r^{(q)}_{\psi,I}(t) \equiv \sup_t \langle P_I(H_\omega)\psi_t, \ |X|^q
P_I(H_\omega)\psi_t\rangle\ <\infty.
$$
Here $\psi_t=e^{-iH_\omega t} \psi$, and $P_I(H_\omega)$ is the spectral 
projector 
of
$H_\omega$ onto the interval $I$.
\end{de}

Our results are the following:

\begin{thm}
\label{thdimer} Let $(H_\omega)_{\omega\in\Omega}$ be as in (\ref{eq:dimer}) and
$V\in]0,1]\backslash\{1/\sqrt{2}\}$. Then, with probability $1$ in $\omega$
 the Lyapounov exponent 
$$
 \gamma(E)=\bd \lim_{x\rightarrow\infty} \ed
\frac{1}{|x|} \ln \left\|T^E_{V_\omega(x)} T^E_{V_\omega(x-1)} \cdots 
T^E_{V_\omega(1)}\right\|
$$
exists, is independent of $\omega$, and :

\vskip1mm
\noindent (i) $\gamma(E=\pm V)=0$ and $\gamma(E\neq\pm V)>0$;

\noindent  (ii) $H_\omega$ has pure point spectrum;

\noindent (iii) Let $\eps>0$ and let $I$ be a  compact energy interval  $I\subset
\sigma(H_\omega)=[-V-2, V+2]$  with $\pm V \not\in I$.  Then, for all
 $0<\gamma<\gamma(I)\equiv \inf\{\gamma(E), E\in I\}$ there exists a constant
$C(\omega,\eps,\gamma)$ and, for each eigenfunction
$\phin$ with energy $\En\in I$, a ``center" $\x\in\Z$, such that  
\begin{equation}
\forall x\in\Z,\;\;|\phin(x)|
  \leq C(\omega,\eps,\gamma) e^{|\x|^\eps} e^{-\gamma |x-\x|};\label{A}
\end{equation}

\noindent Moreover if $\psi$ decays exponentially with mass $\theta>0$ and if $q>0$, 
there exists a constant $C_{\psi,\omega}(I)$ so that~:
\begin{equation}
\bd \sup_t \ed r^{(q)}_{\psi,I}(t) \;\leq\;C_{\psi,\omega}(I) \quad \PP\,a.s.
\label{B}
\end{equation}
In particular, $H_\omega$ is dynamically localized on $I$.
\end{thm}

{\em Remark}: A careful analysis of Lemma 3.5 and 3.6 of \cite{dynloc} shows that our 
estimate
fails ({\em i.e.} $C_{\psi,\omega}(I)$ grows to infinity) if the distance
between $I$ and the energies $\pm V$ decreases ($\gamma\to 0$): this is of course as it 
should
be if one believes that the observed $t^{3/2}$ does indeed occur.

\medskip

These results are completed by the two following theorems:

\begin{thm}
\label{thprim}
Let $(H_\omega)_{\omega\in\Omega}$ be as in (\ref{eq:dimer}),
$V>1$ and $V\neq\sqrt{2}$. Then, for almost all $\omega$, $\gamma(E)$ exists and 
$\gamma(E)>0$ for all $E$, the 
spectrum is pure point 
and (iii) of Theorem~\ref{thdimer} holds with $I=\sigma(H_\omega)$.
\end{thm}

\begin{thm}
\label{thcrit}
Let $(H_\omega)_{\omega\in\Omega}$ be as in (\ref{eq:dimer}) and
$V=\sqrt{2}/2$ (respectively $V=\sqrt{2}$). 
Then the same conclusions as in Theorem \ref{thdimer} (resp. Theorem 
\ref{thprim}) hold except 
at the energies $E_c= \pm 3/\sqrt{2}$ (resp. $E_c=0$). In particular (iii)
hold for intervals $I$ such that $\pm V,\pm 3/\sqrt{2}\not\in I$ (resp. $0\not\in 
I$). In addition $E_c= \pm 3\sqrt{2}/2$ (resp. $E_c=0$) is a 
critical energy in the sense that $\gamma(E_c)=0$.
\end{thm}

%
%

We shall prove Theorems \ref{thdimer} and \ref{thprim} simultaneously in this section, and 
then, in section \ref{sectcrit}, we prove Theorem \ref{thcrit} which deals with the 
critical couples 
$(V=1/\sqrt{2}, \,E_c= \pm 3/\sqrt{2})$ and $(V=\sqrt{2}, \,E_c=0)$.

\medskip

\noindent {\bf Proof of Theorems \ref{thdimer} and \ref{thprim}:} That
(\ref{A}) implies (\ref{B}) is not too hard to see, and is at any rate shown in
\cite{dynloc}, section 2 (see also \cite{mathieu}). To prove (\ref{A}), it will be 
sufficient 
to show
strict positivity of the Lyapunov exponent. Using Theorem 4.1 of
\cite{108} with the transfer matrix  $T_v^E$, this will indeed imply
 the Wegner estimate, which is the ingredient needed to 
make the multiscale analysis function (see the appendix of
\cite{124}, or \cite{carlac}
\cite{pafi}). As a result, one can apply the proof of Theorem 3.1 in
\cite{dynloc}, or equivalently arguments developed in \cite{mathieu}, to
conclude.

We therefore turn to the proof of (i). We first recall it is well known \cite{boulac} 
\cite{cfks} that thanks to the Furstenberg and Kesten Theorem the Lyapunov exponent 
$\gamma$
is well defined on a set
$\Omega_0$ of full measure, and is independent of $\omega\in\Omega_0$.

%
%

\medskip

Consider first the energy $E=V$. The two possible transfer matrices are
\[ T^V_{-V}=\left( \begin{array}{cc}
                4V^2-1 & -2V \\
                2V     & -1
             \end{array}
      \right)
\mbox{ and }\; T^V_V=-Id.
\]
For $\omega\in\Omega_0$ and $x\in\N$, let $n_x = \sharp \{y\in\Z,\, 1\leq y\leq x,
V_\omega(2y)=-V\}$ (this is the number of times $-V$ is obtained after $x$ trials). 
Using the 
following three simple facts:
\begin{itemize}
\item
$\PP \hbox{ a.s. } \bd {\ n_x \over x} \ed\rightarrow p$;
\item
$\bd \lim_{x\rightarrow +\infty} \left\|(T^V_{-V})^x\right\|^{1/x}  = \rho(T^V_{-V}) 
\ed$,
where $\rho(T^V_{-V})$ denotes the spectral radius of $T^V_{-V}$;
\item
$\rho(T^V_{-V}) = 1$, if $V\in]0,1]$, and $\rho(T^V_{-V}) > 1$ if $V>1$;
\end{itemize}
one immediately obtains that $\gamma(E=V)=0$ if $V\in]0,1]$ and $\gamma(E=V)>0$ if 
$V>1$. 
One proceeds similary for the energy $E=-V$.

%
%

\medskip

We now turn to others energies $E\neq\pm V$, and  prove that $\gamma(E\neq\pm 
V)>0$ for all $E$ belonging to the spectrum of $H_\omega$.
Let  $G$ be the smallest 
closed subgroup of SL$(2,\R)$
generated by the matrices $T^E_{V}$ and $T^E_{-V}$. Recall that there is
a natural action of SL$(2,\R)$ on $P(\R^2)$, the set of
all the directions of $\R^2$. A matrix $T\in G$ is then seen as an homography acting on 
$P(\R^2)$. According to the Furstenberg Theorem (see Theorem
I.4.4 of \cite{boulac}), the conclusion will
follow if $G$ is not compact and if either there is no probability measure on 
$P(\R^2)$ that is
invariant under the action of $G$, or equivalently if the orbit 
$G\cdot\tilde{x}\equiv\{T\cdot\tilde{x}, T\in G\}$ of each direction $\tilde{x}\in 
P(\R^2)$
contains at least three elements (Proposition I.4.3 in \cite{boulac}).

In order to alleviate the notations, let's define $\alpha=E-V$ and $\beta=E+V$.
Note that in the present case $\alpha\neq 0$ and $\beta\neq 0$. We will also 
rename
$T^E_V=T_\alpha$ and $T^E_{-V}=T_\beta$, {\it i.e.}
\[ T_X=\left( \begin{array}{cc}
                X^2-1 & -X \\
                X     & -1
             \end{array}
      \right)
\mbox{ with }\; X=\alpha,\,\beta.
\]
We recall that a matrix $T$ is said to be elliptic if $|{\rm tr}\,T|<2$, parabolic 
if 
$|{\rm tr}\,T|=2$ and hyperbolic if $|{\rm tr}\,T|>2$. The proof is reduced to the study 
of 
three cases: a) both the matrices $T_\alpha$ and $T_\beta$ are elliptic; 
b) $T_\alpha$ is parabolic; c) $T_\alpha$ is hyperbolic. These clearly cover all the 
possible cases since the problem is symmetric in $\alpha$ and $\beta$. Note that in 
cases b) and c) the group $G$ is clearly not compact.

%
%

\medskip
\noindent
Case a).
Suppose $T_\alpha$ and $T_\beta$ are
both elliptic, {\em i.e.} $|\alpha|,|\beta|\in ]0,2[$. In that case they do not 
commute, since $E\not=V$. Since the commutator 
$T=T_\alpha T_\beta \left(T_\alpha\right)^{-1} \left( T_\beta\right)^{-1}$
of two non-commuting elliptic elements is known to be hyperbolic ($|{\rm
tr}\,T|> 2$) - see the proof of Proposition 2.8 in \cite{iversen} - it follows 
that $G$ is not compact. We will show $G\cdot\tilde{x}$ contains at least three 
points provided $\alpha^2\neq 2$ or $\beta^2\neq 2$. 

To that end, note first that
${\rm tr}T_X^2=X^4-4X^2+2$, so that if $X^2\in]0,4[$ and $X^2\neq 2$, then $T^2_X$ 
is elliptic. Hence, if $\alpha^2\neq 2$ or $\beta^2\neq 2$, then $T_\alpha$ and
$T_\alpha^2$ or $T_\beta$ and $T_\beta^2$ are elliptic. Since elliptic elements 
have no fixed points in $\PP(\R^2)$, it follows easily that for any 
$\tilde{x}\in\PP(\R^2)$, $G\cdot\tilde{x}$ contains at least the three points 
$\tilde{x}$, $T_X\cdot \tilde{x}$, $T_X^2\cdot\tilde{x}$, with $X=\alpha$ or $\beta$.

If, on the other hand, $\alpha^2= 2$ and $\beta^2= 2$, then $E=0$ and
$V=\sqrt{2}$, which is one of the two critical couple described 
in Theorem \ref{thcrit}, and to be dealt with in section \ref{sectcrit}.

%
%

\medskip
\noindent
Case b).
Suppose now that $T_\alpha$ is parabolic, {\em i.e.} $|\alpha|=2$. We treat the 
case $\alpha=2$ (the case $\alpha=-2$ is similar). The
eigenvector of $T_\alpha$ is then given by $(1,1)$. 
Denoting by $e_2$ the orthogonal vector $(1,-1)$, the matrix $T_\alpha$ in the 
basis
$(e_1,e_2)$ can be written
$$
\left( \begin{array}{cc}
                1 & 4 \\
                0 & 1
             \end{array}
      \right),
\mbox{ and so }\;
\left( \begin{array}{cc}
                1 & 4 \\
                0 & 1
             \end{array}\right)^n=\left( \begin{array}{cc}
                                1 & 4n \\
                                0 & 1
                              \end{array}
                        \right).
$$
Taking a vector $x=x_1e_1+x_2e_2$, and writing $\tilde{x}$ for
its direction ({\em i.e.} its projection onto
$P(\R^2)$), one concludes that 
$\lim_{n\rightarrow\infty}T_\alpha^n\cdot\tilde{x}=\tilde{e}_1$
(where $T_\alpha^n$ is seen
here as a homography of $P^2(\R)$). But now, if $m$ is a probability measure that 
is invariant
under the action of $G$, and if $f\in C_0^\infty(P(\R^2))$, using a Lebesgue 
dominated
argument, one has
$$
f(\tilde{e}_1)= \bd \lim_{n\longrightarrow\infty} \ed \int f(T_\alpha^n \cdot
\tilde{x})dm(\tilde{x})
=\langle m,f\rangle.
$$
This means that $m=\delta_{\tilde{e}_1}$. But now one uses the second matrix 
$T_\beta$: it does
not leave invariant the direction $\tilde{e}_1$ except for $\beta=0$ or 
$\beta=2=\alpha$ (simple check),
which is excluded since the first condition yields $E=-V$ and the second one 
$V=0$. Thus we proved there is no invariant measure in
case b).

%
%

\medskip
\noindent
Case c).
Suppose now that $T_\alpha$ is hyperbolic ($|\alpha|>2$). It is clearly sufficient 
to study the orbit of the eigendirections of $T_\alpha$, namely
$e_\eps=(\alpha+\eps\sqrt{\alpha^2-4},2)$, $\eps=\pm 1$. Note that $T_\alpha$ and these 
$T_\beta$ cannot have eigenvectors in common, since it is easy to show that it would imply
$\alpha=\beta$ (and $V=0$).
Now, if $T_\beta$ is hyperbolic then it is clear that the orbit of $e_\eps$ is infinite.
If $T_\beta$ is parabolic then we are again in case b). Finally, if $T_\beta$ is elliptic 
then let's consider $\tilde{X}\equiv T_\beta\widetilde{e_\eps}$. If $\tilde{X}\neq 
\widetilde{e}_{-\eps}$ then $\tilde{X}$ cannot belong to the eigendirections of $T_\alpha$ 
and its orbit is then
infinite.

Hence, the only case we still need to
consider is the case where $T_\beta$ is elliptic and exchanges these two
directions (the orbit of these elements would then have cardinal 2). In that case 
$T_\beta e_\eps$ and $e_{-\eps}$, $\eps=\pm1$, have
the same directions, and simple calculations lead to the two equations 
$$
(\beta^2-1)(\alpha+\eps\sqrt{\alpha^2-4})=4\beta-(\alpha-\eps\sqrt{\alpha^2-4}),
\quad \eps=\pm 1.
$$
It trivially implies $\beta^2=2$ and $\alpha=2\beta$, which means $V=\sqrt{2}/2$ and
$E=-3\sqrt{2}/2$. The symmetric case where one assumes that $T_\beta$ is hyperbolic leads 
naturally to $\alpha^2=2$ and $\beta=2\alpha$, which means this time $V=\sqrt{2}/2$ and
$E=3\sqrt{2}/2$. Since, in Theorem \ref{thdimer} we have supposed 
$V\neq\sqrt{2}/2$, the proof is complete.

%
%
\section{New critical cases}
\label{sectcrit}

We now consider the two special cases which haven't been studied in the previous section 
and that are dealt with in Theorem~\ref{thcrit}, that is $(V=1/\sqrt{2}, E_c=\pm 3/\sqrt{2})$ 
and 
$(V=\sqrt{2}, E_c=0)$.

\vskip2mm\noindent
{\bf Proof of Theorem \ref{thcrit}:}

It clearly follows from the previous proof that the only thing that remains to be 
proven is that the Lyapunov exponent is zero at the critical energies $E_c$. Note 
first that in all cases $E_c$ belongs to the spectrum of $H_\omega$ almost surely since 
$d(-3\sqrt{2}/2,-1/\sqrt{2})=d(3\sqrt{2}/2,1/\sqrt{2})=d(0,\pm \sqrt{2})=\sqrt{2}<2$.

\medskip
We first deal with the critical case $(V=1/\sqrt{2}, E_c=\pm 3/\sqrt{2})$. The second one 
will then be easier to treat.
\medskip

\noindent\underline{$(V=1/\sqrt{2}, E_c= \pm 3/\sqrt{2})$}

\medskip

Clearly it is enough to restrict ourselves to the case $E_c= - 3/\sqrt{2}$.
Using the notations and the results of the previous proof, we thus have, in the 
present case, $\beta^2=2$ and $\alpha=2\beta$. The 
eigenvectors of $T_\alpha$ are then given by $(\beta+\eps,1)$, $\eps=\pm 1$, and looking 
at the 
matrices in the basis of these two vectors we are reduced to considering products of 
matrices of the following two types:
\[ \left( \begin{array}{cc}
                \lambda_1 & 0 \\
                     0    & \lambda_2
             \end{array}
      \right)
\mbox{ and }\;
     \left( \begin{array}{cc}
                0 & 1-\beta \\
                     1+\beta    & 0
             \end{array}
      \right),
\]
with $\lambda_1\lambda_2=1$, $\lambda_1>1$, and $(1-\beta)(1+\beta)=-1$. With 
some abuse of notation, we will 
again denote those two matrices by $T_\alpha$ and $T_\beta$.

To prove that $\gamma(V=\sqrt{2}/2, E=-3\sqrt{2}/2)=0$, one has to analyse, 
roughly  speaking, the behaviour of large products of matrices $T_\alpha$ and 
$T_\beta$. 
While the matrices $T_\alpha$ contribute to the growth of the norm of such a 
product, the $T_\beta$ not only do not contribute (being a rotation) but in fact 
``destroy" 
this growth. Indeed one checks
\begin{eqnarray}
T_\beta\,T_\alpha^n\, T_\beta & = &
      - \left( \begin{array}{cc}
                 \lambda_2^n & 0 \\
                     0    & \lambda_1^n
             \end{array}
        \right) \nonumber \\
\mbox{ and } \quad
T_\beta\,T_\alpha^{n_2}\,T_\beta\,T_\alpha^{n_1} & = &
-\left( \begin{array}{cc}
                \lambda_1^{n_1-n_2} & 0 \\
                     0    & \lambda_2^{n_1-n_2}
             \end{array}
      \right), \label{eq:a}
\end{eqnarray}
since $\lambda_1\lambda_2=1$ and $(1-\beta)(1+\beta)=-1$. Noting that $T_\beta^2=-Id$, a 
product of 
factors $T_\alpha$ and $T_\beta$ is, up to a sign, a succession of 
$T_\alpha^{n_i}$ and 
$T_\beta$. One then easily understands, from (\ref{eq:a}), that the norm of a product 
$T_n\dots T_1$ can not grow fast enough to ensure the positivity of the Lyapunov 
exponent.
This is exactly what we show below. 

\medskip
We will see a product $T_n\dots T_1$ as a sequence of $m(n)$ steps, where $m(n)$ is the 
number the matrices $T_\alpha$ contained in the chain $T_n\dots T_1$; in other words a 
``step" means that one matrix $T_\alpha$ has been met. So each step is a product of 
matrices 
of the form $T_\beta^{j_i} T_\alpha$. So $T_n\dots T_1$ will be written 
$\prod_{i=0}^{m(n)-1} \left(T_\beta^{j_i} T_\alpha\right)$. Looking at (\ref{eq:zzz}), it is 
clear that without loss of generality on can suppose $T_1=T_\alpha$. Clearly, depending on the 
parity of $j_i$, the $i^{\rm th}$ step will 
contribute or not to the growth (in norm) of the total product $T_n\dots T_1$.

More precisely, in order to study the product of elements of the form $T_\beta^{j_i} 
T_\alpha $, we define two sequences $u_k$ and $V_k$ such that, after $k$ steps,
\begin{equation}
\prod_{i=0}^{k-1} \left(T_\beta^{j_i} T_\alpha\right)= \pm T_\beta^{u_k} T_\alpha^{V_k},
\label{eq:uV}
\end{equation}
with $u_k\in\{0,1\}$. This is clearly always possible using relations (\ref{eq:a}) and 
$T_\beta^2=-Id$. Now it is easy to obtain 
recurrence relations for $\eps(u_k)=(-1)^{u_k}$ and $V_k$:
\begin{eqnarray*}
 T_\beta^{u_{k+1}} T_\alpha^{V_{k+1}}
& = & \pm\left(T_\beta^{j_k} T_\alpha\right) T_\beta^{u_k} T_\alpha^{V_k}\\
& = & \pm\left\{ \begin{array}{l}
                T_\beta^{j_k} T_\alpha^{V_k+1} \mbox{ if } \eps(u_k)=1\\
                T_\beta^{j_k+1} T_\alpha^{V_k-1} \mbox{ if } \eps(u_k)=-1.
              \end{array}   
      \right.
\end{eqnarray*}
And this leads to
\begin{equation}
\left\{ \begin{array}{l}
            V_{k+1}=V_k+\eps(u_k)\\
            \eps(u_{k+1})=\eps(j_k)\eps(u_k).
\end{array}   
      \right.
 \label{eq:rec}
\end{equation}
Then define $\eps_k=\eps(j_{k-1})$ and $U_k=\eps(u_k)=\eps_k\cdots\eps_1$, for $k\geq 
1$.
So $\eps_k$ is a sequence of independent and identically distributed 
random variables  taking the two values $\pm 1$, and such that $\eps_k=+1$ if one meets 
an even(possibly 
zero) number of $T_\beta$ between the $(k-1)^{\rm th}$ and the $k^{\rm th}$ matrix 
$T_\alpha$, and  $\eps_k=-1$ if not. Let's recall that $\PP(T_\alpha)=1-p$ and 
$\PP(T_\beta)=p$. So one has
$$
P(\eps_k=1)=(1-p)(1+p^2+...)=\frac{1-p}{1-p^2}={1\over p+1}
$$
and
$$
P(\eps_k=-1)=(1-p)(p+p^3+...)=\frac{p(1-p)}{1-p^2}={p\over p+1}.
$$
Moreover one checks $\E(\eps_k)=(1-p)/(1+p)\in ]0,1[$ since $p\in ]0,1[$.
Finally let us rewrite equations (\ref{eq:rec}) as
$$
U_k=\prod_{i=1}^k \eps_i \quad
\mbox{ and } \quad
V_m=\sum_{k=1}^m U_k.
$$
To understand how these random sequences behave, note that 
$U_{k+1}=\eps_{k+1}U_k\in\{-1,1\}$. So, if $\eps_{k+1}=1$ then 
$U_{k+1}+U_k=\pm 2$, but if $\eps_{k+1}=-1$, then $U_{k+1}+U_k=0$. As a result
looking at the sum $V_m$, $U_{k+1}$ destroys in the latter case the term before, and 
does not 
contribute to the growth of $V_m$.

Note that one can prove from (\ref{eq:rec}) that
$$
\PP(U_k=1)={1\over 2}\left( 1+\left( {1-p \over 1+p}\right)^k\right)
\quad {\rm and} \quad
\PP(U_k=-1)={1\over 2}\left( 1-\left( {1-p \over 1+p}\right)^k\right).
$$

By construction $V_m$ in turn is closely related to the 
exponential 
growth of the product $T_n\dots T_1$, as one can see from the following formula:
\begin{eqnarray}
\ln\|T_n\dots T_1\| & = & \ln \left\|\prod_{i=0}^{m(n)-1} T_\beta^{j_i}T_\alpha \right\|
=\ln \left\|T_\beta^{u_{m(n)}}T_\alpha^{V_{m(n)}} \right\|  \nonumber\\
 & \leq & |V_{m(n)}|\ln\lambda_1 + \ln\|T_\beta\|.  \label{eq:zzz}
\end{eqnarray}
Since, by the Furstenberg and Kesten Theorem \cite{boulac} \cite{cfks}, 
$\gamma$ exists almost surely and is constant, and since $\bd {1\over n} \ed 
\ln\|T_n\dots T_1\|\leq 
\max(\|T_\alpha\|,\|T_\beta\|)$, the Lebesgue dominated convergence Theorem gives
$$
\gamma =\E\left(\lim_{n\to\infty}\ln\|T_n\dots T_1\|/n\right)= \ln\lambda_1 
\lim_{n\to\infty} 
\E\left(|V_{m(n)}|/n\right).
$$
It remains to evaluate the latter limit. Computing $V^2_{m(n)}$ one obtains that
\begin{eqnarray}
V_{m(n)}^2 & = & \sum_{k=1}^{m(n)} U_k^2 +2\sum_{1\leq k < l \leq m(n)} U_k U_l 
\nonumber 
\\
           & = & m(n)+2\sum_{1\leq k < l \leq m(n)} \eps_{k+1}\cdots \eps_l, 
\label{eq:V2}
\end{eqnarray}
since $\eps_k^2=1$. Moreover, using the independence of the $\eps_i$, one has 
$\E(U_k U_l)=\E(\eps_1)^{|k-l|}$; but $m(n)$ does also depend on $\omega$ (write 
$m(n,\omega)$). So one needs some control on how  $m(n,\omega)$ depends on $\omega$. 
This 
is 
provided 
by the following lemma, which just recalls well-known results about Bernoulli random 
variables ({\em 
e.g.} \cite{loeve}). 

\begin{lem}
Let $m(n,\omega)$ be the number of $T_\alpha$ contained in the product 
$T_n(\omega)\cdots 
T_1(\omega)$. One has
$$
\overline{m_n}\equiv\E(m(n,\omega))=(1-p)n,
$$
and
$$
{\rm Var}(m(n,\omega))=\E\left[(m(n,\omega)-\overline{m_n})^2\right] = p(1-p)n.
$$
\end{lem}

An immediate consequence of this lemma is that
\begin{equation}
\E\left[\left(m(n,\omega)-[\overline{m_n}]\right)^2\right] \sim p(1-p)n \quad \mbox{ as } 
n\to\infty, 
\label{eq:sim}
\end{equation}
where $[\overline{m_n}]$ denotes the integer part of $\overline{m_n}$.
Then the result follows from
\begin{eqnarray*}
\lefteqn{\E(|V_{m(n)}|/n)} \\
& \leq & \frac{1}{n}\E\left(\left|V_{m(n,\omega)}- 
V_{[\overline{m_n}]}\right|\right)+ 
\frac{1}{n}\E\left(\left|V_{[\overline{m_n}]}\right|\right) \\
 & \leq & \frac{1}{n} \sqrt{\E\left(\left(m(n,\omega)-[\overline{m_n}]\right)^2\right)}
         + \frac{1}{n}\sqrt{[\overline{m_n}]+2\sum_{1\leq k < l \leq [\overline{m_n}]}        
             \E(\eps_{1})^{|k-l|}} \\
 & \leq &  \frac{C}{\sqrt{n}},
\end{eqnarray*}
 for some constant $C>0$, where we used, successively, the Cauchy-Schwartz inequality, 
relations 
(\ref{eq:V2}) and (\ref{eq:sim}), and 
the facts that
$\E(\eps_1)<1$ and $[\overline{m}]\leq n$. In conclusion it follows that 
$\E(\gamma(E=-3\sqrt{2}/2))=0$.

\medskip
We now turn to the second special case.
\medskip

\noindent\underline{$(V=\sqrt{2}, E_c=0)$}

\medskip
So $\alpha=-\beta=\pm \sqrt{2}$ and let us recall that 
\begin{equation}
T_\alpha^2=T_\beta^2=-Id. \label{eq:id}
\end{equation}
We shall follow the idea of the previous case, but the arguments are much simpler. Regrouping 
all the powers of $T_\alpha$ and $T_\beta$ that appear in the product of the 
$n$ first matrices $T_n\cdots T_1$
and taking (\ref{eq:id}) into account, the product $T_n\cdots T_1$ can be 
reduced (essentially) to some power $V_n$ of the matrix $T_\alpha T_\beta$ which is 
hyperbolic. This would then lead to a strictly positive Lyapunov exponent (since the 
spectral radius of $T_\alpha T_\beta$ is strictly greater than 1) if $V_n$ and $n$ had 
the same order, which is however not the case.

Let us consider groups of two matrices in the product $T_n\cdots T_1$. Then it is easy to see 
that one can define a sequence $V_k$ with $V_0=0$ and
$$
T_{2k}\cdots T_1=(T_\alpha T_\beta)^{V_k}.
$$
Depending on the values of $T_{2k+2}$ and $T_{2k+1}$, and noting that $(T_\alpha 
T_\beta)^{-1}= T_\beta T_\alpha$, one has
\begin{equation}
\begin{array}{c}
\PP(V_{k+1}=V_k+1)=p(1-p)=\PP(V_{k+1}=V_k-1),\\
\quad \\
{\rm and} \quad \PP(V_{k+1}=V_k)=p^2+(1-p)^2.
\end{array} \label{eq:loiU}
\end{equation}
This situation is different from the previous one where the way the value of $V_k$ changed 
(between the 
$k^{\rm th}$ and $(k+1)^{\rm th}$ steps) was 
depending on what happened before.
So let us define $U_k=V_{k+1}-V_k$. It is (unlike before) an i.i.d random sequence the law of 
which is given by (\ref{eq:loiU}).
One easily computes $\E(U_k)=0$ and $\E(U_k^2)=2p(1-p)$. It is then immediate that
\begin{eqnarray*}
\E(V_n^2) & = & \sum_{k=1}^n \E(U_k^2) +2\sum_{1\leq k < l \leq n} \E(U_k U_l)
\nonumber 
\\
           & = & 2np(1-p), 
\end{eqnarray*}
since $\E(U_k U_l)=\E(U_k) \E(U_l)=0$ for $l\neq k$. The result then follows in the same 
way as previously. \fin

\medskip

{\em Remark:} the situation is, to our opinion, 
different 
from the one we met with the critical energies $E=\pm V$. It is worth to notice that if 
$E=\pm V$ then $\bd \lim_{n\to+\infty} {1\over n^{\nu}} \ed \ln\|T_n\dots T_1\|=0$ for 
all $\nu>0$, since it is easy to see that in this case $\|T_n\dots T_1\|$ is bounded 
independently of $n$. 
We conjecture that this is not the case at the critical couples $(V=1/\sqrt{2}, 
E_c=\pm 3/\sqrt{2})$ and $(V=\sqrt{2}, E_c=0)$, where for $\nu<1/2$ the limit is probably 
infinite (and zero for $\nu>1/2$). If so it is reasonable to think that the eigenfunctions 
with energy $E$ close to $E_c$ should decay sub-exponentially (semi-uniformly) as $\exp 
-\gamma(E)n^\nu$ ($\nu<1/2$). This would still imply dynamical localization even on a 
spectral interval containing the critical energy $E_c$.

\medskip
{\bf Acknowledgement:} It is a pleasure for the authors to thank Gian-Michele 
Graf warmly for a careful reading of a previous version of this paper.

%
%

\end{document}